\documentclass[pre, preprint,floatfix]{revtex4}
\usepackage{graphicx}
\renewcommand{\r}{\right}
\renewcommand{\l}{\left}

\begin{document}

\title{Global Snapshot of Protein Interaction Network -- A Percolation Based Approach}

\author{Chen-Shan Chin}

\affiliation{Department of Biochemistry and Biophysics, University of
  California, San Francisco, 94143, CA, USA}

\email{cschin@genome.ucsf.edu}

\author{Manoj Pratim Samanta}

\affiliation{NASA Advanced Supercomputing Division, NASA Ames Research Center,
  Moffet Field, 94035, CA, USA}

\email{msamanta@nas.nasa.gov}

\date{\today}

\begin{abstract}
  In this paper, we study the large-scale protein interaction network
  of yeast utilizing a stochastic method based upon percolation of
  random graphs.  In order to find the global features of
  connectivities in the network, we introduce numerical measures that
  quantify (1) how strongly a protein ties with the other parts of the
  network and (2) how significantly an interaction contributes to the
  integrity of the network.  Our study shows that the distribution of
  essential proteins is distinct from the background in terms of
  global connectivities.  This observation highlights a fundamental
  difference between the essential and the non-essential proteins in
  the network.  Furthermore, we find that the interaction data
  obtained from different experimental methods such as
  immunoprecipitation and two-hybrid techniques possess different
  characteristics.  We discuss the biological implications of these
  observations.
\end{abstract}

\maketitle

\section{Introduction}

Recent availability of a large amount of data from high-throughput
experiments~\cite{Zhu2,Uetz,Ito,Gavin,Ho} has brought about a
fundamental change in the way we study biological systems. Unlike the
traditional methods which relied on probing a single or a few proteins
to identify important pathways, it is now becoming possible to
describe larger functional `modules'~\cite{Hartwell} and even the
global properties of the entire
proteome~\cite{Jeong,Maslov,Mering,Bader}.  Researchers are attempting
to connect large-scale protein interaction data with information from
phenotype studies~\cite{Jeong,Maslov}.  In one such analysis of data
from yeast, Jeong {\it et al.}  observed the connectivities of
individual proteins in the network to closely follow a power-law
distribution.  Similar to other power-law networks, positive
correlation existed between a protein's inviability and its
connectivity~\cite{Jeong}.  In another study, Maslov {\it et al.}
observed interesting patterns in the distribution of the links between
the nearest neighbors in the network and postulated that such patterns
give rise to the specificity and the robustness of the
network~\cite{Maslov}.

One of the shortcomings of the previous approaches is that they drew
conclusions about the global nature of the network from its local
connectivity properties. It is unclear whether such local studies
based on individual nodes or nearest neighbors fully capture the
global picture of the network. For example, some essential proteins,
namely, those for which null mutants produce inviable
strains~\cite{YeastDel}, may have few numbers of direct links but
still take important roles in the network through the proteins to
which they are connected.  Such proteins would not be correctly
identified by just counting the number of links as in
Ref.~\cite{Jeong}.  To properly recognize such cases, it is necessary
to go beyond the nearest neighbor links.  However, it is not clear
that the techniques mentioned above can easily be extended to answer
such questions.

In this paper, we introduce a stochastic method inspired by the
percolation model in statistical mechanics\cite{percolation} that
overcomes the shortcomings of the previous approaches.  This method
allows us to define a quantity that measures the correlation between
any two nodes in the network, taking the topology of the entire
network into account.  Biologically, such correlations describe the
direct and indirect influences of one protein on another through the
protein interaction network.  If such correlations indeed carry
biological significance, we expect the essential proteins to be highly
correlated, in general, with the rest of the network.  One of our main
results is that most essential proteins do possess higher correlations
between themselves and the rest of the network.  This is consistent
with previous results~\cite{Jeong}, because in the first order, the
correlations computed by us are proportional to the connectivities of
the proteins. However, we show that it is important to go beyond the
first order. Identifying essential proteins by our method performs
consistently better than just counting links.  Additionally, we
observe that the essential proteins interact more tightly with the
other essential proteins, thus forming a `network core'.  This
directly agrees with large-scale experiments probing protein
networks~\cite{Gavin}.

Based on our method, we can also quantify the relative significance of an
interaction to the integrity of the network. We observe that the
interaction data from different measurement techniques, such as
immunoprecipitation(IP) and the two-hybrid test, give distinct
distributions.  This suggests that various experimental
techniques for probing the protein interaction might explore
different regions of the network.

\section{Method and Materials}
\label{sec:method}

\subsection{Bond-percolation on Graph}
Given any two nodes in a network, the strength of their connectivity
can be estimated in different ways. Some of these measures are local.
For example, we can ask whether any two nodes are directed linked, how
many common neighbors they share~\cite{Samanta}, {\it etc}.  We can
also ask how local properties of a node, such as the degree of links,
associate with its function and its importance in the
network\cite{Jeong}.  Furthermore, information about the correlations
between nodes involving nonlocal properties, such as the length of the
shortest path and clustering structures, will enable us to uncover
hidden features buried within the massive data. Here, we present a
generic approach that extracts useful information about a node beyond
its local connections.

Correlations between two nodes may come from other numerous short
paths rather than just the shortest path.  A reasonable estimate of
correlation should take into account the number and lengths of
different paths between two nodes.  One possible way to estimate such
correlation between two nodes is to repeatedly remove some fraction
$q$ of the links in the network chosen randomly and check whether they
still remain connected.  Their probability remaining
connected is proportional to the number of short paths between them
and inversely proportional the length of those paths.  This
probability provides a good measurement of the correlation between two
nodes that includes the information regarding the non-local topology
of the network.  The described process of finding the correlation
between two nodes in a network is equivalent to the bond-percolation
model in statistical mechanics\cite{percolation}.

Mathematically, a network is treated in the language of graph theory,
where a node is denoted as a vertex and a link as an edge.  Given a
graph $G$ with vertices $V$ and edges $E$, a percolation configuration
is realized as follows.  Each edge $e_{ij}$ linking vertices $i$ and
$j$ is assigned a random number $p_{ij}$ distributed uniformly from 0
to 1.  If this random number is greater than $p = 1 - q$, a given
percolation probability, then the edge is eliminated from the original
graph.  The final graph $G^\prime$ consists of the edge set $E^\prime
= E - \bar{E}$, where $\bar{E}$ is the set of edges that $p_{ij} > p$
and $E^\prime$ consists those edges with $p_{ij} < p$.  Assuming that
$G$ is connected, the reduced graph $G^\prime$ may or may not remain a
single connected component depending on $p$.

\subsection{Susceptibility}
The first step in applying the algorithm is to determine the appropriate
value of the probability $p$. If $p$ is near one, then we only produce
totally connected graphs.  If $p$ is too close to zero, then the network
is split into individual vertices and small clusters. An intermediate value of 
$p$ provides information about the non-local properties of the network.

The degree of fragmentation in the graph $G^\prime$ can be quantified
by the order parameter $m(p)$, the ratio of the largest connected
component to the total graph size.  It is defined as $m(p) = N_{\rm
  max}/|V|$, where $N_{\rm max}$ is the number of vertices of the
largest connected component and $|V|$ is the total number of vertices.
For a connected graph $G$, $m(p)$ varies from $1/|V|$ to 1 as $p$
changes from 0 to 1.  Here, $m$ is a stochastic variable, whose
fluctuation is defined by
\begin{equation}
  \chi(p) = \langle (m - \langle m \rangle)^2 \rangle^{\frac{1}{2}}
\end{equation}
The brackets denote the ensemble average, which is the average over
many different realizations of $G^\prime$.  The curve of $\chi(p)$
reveals certain aspects of the graph topology.  For example, if $G$ is
a regular two dimensional square lattice, then $\chi$ diverges with a
power law behavior as a function of $p-p_{\rm c}$, for $p_{\rm
  c}=1/2$.  For other types of regular lattices, like triangular
lattices or higher dimensional lattices, $p_{\rm c}$ and/or the power
law exponent also change.  A maximum in $\chi(p)$ occurs at the
transition point $p_{\rm c}$, indicating a phase transition and
critical behavior\cite{percolation}.  At this critical point, the
distribution of the sizes of the connected clusters decay as a power
law. Chosing a value of $p$ near this critical value, we get the most
non-local information regarding the network.

\subsection{Correlations and the definition of $v_i$}
Whether two arbitrary vertices $i$ and $j$ remain connected in
$G^\prime$ can provide more detailed information about $G$.  If two
vertices retain their connection, it means that there exist paths in
$E^\prime$ from vertex $i$ to vertex $j$.  Define $\delta_{ij}$ as
function of a pair of vertices $i$ and $j$ such that $\delta_{ij} = 1$
if vertices $i$ and $j$ are connected, and $\delta_{ij} = 0$
otherwise.  The percolation correlation $c_{ij}$ is then defined as the
ensemble average of $\delta_{ij}$,
\begin{equation}
  c_{ij} = \langle \delta_{ij} \rangle.
\end{equation}

With knowledge of the $c_{ij}$, we are equipped to
measure how strongly a vertex $i$ links to the rest of the network
counting both direct and indirect connections to vertex $i$. 
We define the quantity $v_i$ for vertex $i$,
\begin{equation}
  v_i = \frac{1}{|V|} \sum_{j \in V} c_{ij}
\end{equation}
This value is sensitive not only to the linking degree at each vertex
but also to higher order connections between a vertex and the rest of
the random graph.  Thus, $v_i$ effectively ranks the importance of a
vertex in the graph. Intuitively, $v_i$ may be interpreted as the
fraction of other vertices to which vertex $i$ remains linked, if each
edge is broken with probability $q = 1 - p$ in the graph $G$.  In
Fig.~\ref{fig:smallnet}, we show the descending ranking order of the
$v_i$'s for a small graph.

\subsection{The definition of $\beta_{ij}$}
\begin{figure*}[htbp]
  \includegraphics[width=6in]{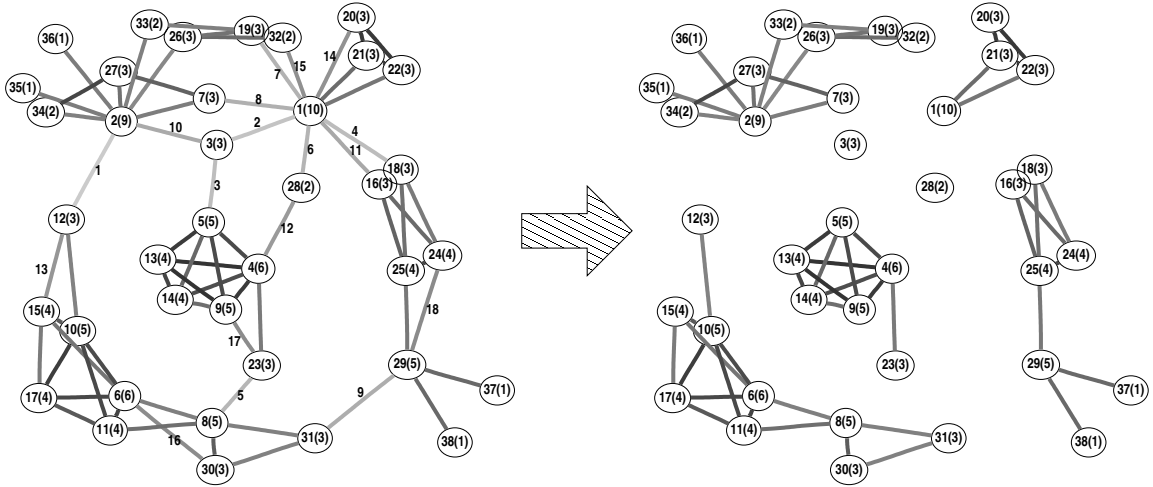}  
  \caption{We applied our algorithm with $p=0.43$ on a small graph.  
    The vertices are indexed in the descending order of $v$ and the
    parenthesized numbers indicate the degree of connection.  Some
    vertices, like vertex 3, have few neighbors but are out-ranked in
    terms of $v_i$ to other vertices with more neighbors.  Vertices
    with equivalent degree of connectivity might be ranked very
    differently because they have differing number of next nearest
    neighbors.  The edges having largest eighteen $\beta_{ij}$ shown
    in gray and are ranked.  If we remove these edges, the graph is
    severed into several compact subgraphs.  The edges carrying
    largest $\beta_{ij}$ tend to link different large components.  The
    edges within a clique, like vertices 5,4,9,13, and 14, have the
    smallest $\beta_{ij}$.}
  \label{fig:smallnet}
\end{figure*}

Using a similar idea, we can define a quantity that allows us to check
the influence of an edge on the graph integrity.  The elimination of
some edges may fundamentally change the connectivity properties
whereas the graph topology may be relatively unchanged against the deletion
of others.  For example, for a small fully connected subgraph, termed
a clique, removal of a certain number of edges between the vertices of
the subgraph tends not to separate the graph into disconnected pieces.
Individual links in the subgraph do not play crucial roles in
supporting the integrity of the subgraph and the whole graph.  We
define the quantity $\beta_{ij}$ to monitor the importance of edge
$e_{ij}$ to the integrity of the graph,
\begin{widetext}
\begin{equation}
  \beta_{ij} = \frac{1}{|V|^2} \sum_{l,m\in V}
  \l(c_{lm}\l(G^\prime \cup \{e_{ij}\}\r) - c_{lm}\l(G^\prime \setminus \{e_{ij}\}\r)\r).
\end{equation} 
\end{widetext}
The first term in the summation is correlation $c_{lm}$ measured by adding
$e_{ij}$ in $G^\prime$ independent of $p_{ij}$ and $p$.  The second
term in $c_{lm}$ measured by removing $e_{ij}$ in $G^\prime$.  The
difference in measurement of $c_{lm}$ under the presence or absence of
edge $e_{ij}$ allows us to distinguish edges.  For example, if
$e_{ij}$ bridges two clusters, then $\beta_{ij}$ will be elevated
(note the edges 1, 2 and 3 in Fig.~\ref{fig:smallnet}).  Suppose edge
$e_{ij}$ connects two disjoint connected components $A$ and $B$ with
sizes $n_{\rm A}$ and $n_{\rm B}$.  Then, in a realization of
$G^\prime$, the contribution to $\beta_{ij}$ is the difference between
$\sum_{l,m\in A\cup B} \delta_{lm} = |n_A+n_B|^2$ and $\sum_{l,m\in
  A} \delta_{lm} + \sum_{l,m\in B} \delta_{lm} = |n_A|^2+|n_B|^2$.
Namely, the contribution to $\beta_{ij}$ is proportional to $n_{\rm
  A}n_{\rm B}$.  However, if $e_{ij}$ is embedded within a connected
component such that adding or removing $e_{ij}$ does not perturb the
component's connectivity, then $e_{ij}$ is redundant and does not
contribute to $\beta_{ij}$.  With this interpretation, $\beta_{ij}$
measures how well $e_{ij}$ succeeds in connecting differing big
components or modules.

\subsection{Protein interaction data}
Here, we apply the described method on the yeast protein interaction
data taken from the Database of Interacting
Proteins(DIP)~\cite{Deane}.  The dataset contains 14871 interactions
between 4692 proteins\footnote{We used the files ``yeast20020901.lst''
  and ``dip20020616.xin'', downloaded from DIP database
  ({\tt http://dip.doe-mbi.ucla.edu/}).} and includes interactions measured
by different experimental methods.  We treat the interaction network
as an undirected graph, with the proteins as vertices. If two proteins
are interaction partners in the dataset, the corresponding vertices
are joined by an edge.

\section{Results and Discussions}
\label{sec:DIP}

\subsection{Determination of $p$}
As a first step in applying this stochastic method on the protein
interaction network, we need to determine the appropriate value of $p$. If
$p$ is near one, then we will only produce totally connected graphs.
If $p$ is too close to zero, then we will only obtain information
about small clusters. Some intermediate value of $p$ will give us
global properties of the network.

\begin{figure}[htbp]
  \includegraphics[width=3in]{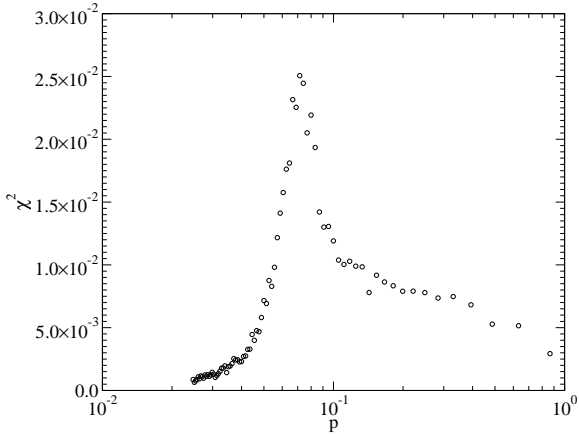}
  \caption{Susceptibility curve of the parameter $m$.  The curve
    peaks at $p=0.07$, where the fluctuations of $m$ are greatest.}
  \label{fig:sus}
\end{figure}

In order to determine the proper value of $p$, we need to compute the
curve $\chi(p)$.  Such a curve for the DIP data is shown in
Fig.~\ref{fig:sus}.  The curve peaks at about $p=0.07$, where the size
fluctuations of the largest cluster are maximal.  Most realizations of
the percolation graph $G^\prime$ in the neighborhood of this peak
yield sparse but still predominantly connected graphs.  Accordingly,
computing $v_i$ and $\beta_{ij}$ around this peak in $\chi(p)$ avoids
the finite size effect at smaller $p$ and loss of resolutions at
larger $p$.

\subsection{Distribution of $v_i$} 
\begin{figure}[htbp]
  \includegraphics[width=3in]{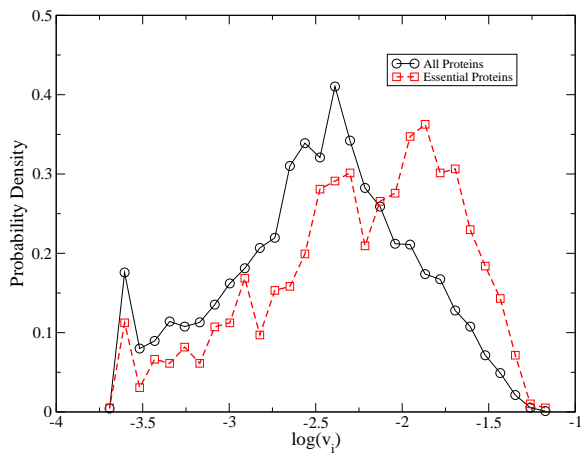}
  \caption{Histogram of $\log(v_i)$.  The distribution of $v_i$ for
    essential proteins is skewed toward larger $v$.}
  \label{fig:hist_vi}
\end{figure}

\begin{figure}[htbp]
  \includegraphics[width=3in]{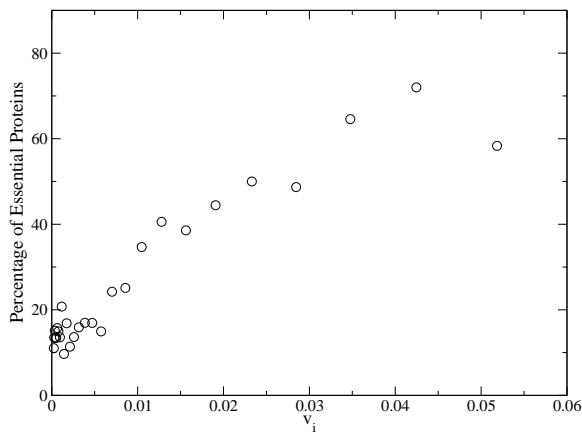}
  \caption{The percentage of proteins which are essential as a
    function of $v_i$. }
  \label{fig:corr-ess-v}
\end{figure}

We gathered our data from $10^5$ realizations of the graph at $p =
0.07$.  The distribution of $\log(v_i)$ for the protein interaction
network is shown in Fig.~\ref{fig:hist_vi}. We also report the
distributions of a subset composing only the essential
proteins\footnote{We got the list of essential proteins from the
  Saccharomyces Genome Deletion Project~\cite{YeastDel}
  ({\tt http://yeastdeletion.stanford.edu/}).}.
The distribution of $v_i$ for essential proteins significantly differs
from the background distribution and is biased toward greater $v_i$.
A protein with a greater $v_i$ ties to the network more strongly than
a protein possessing a smaller $v_i$.  Therefore, we would predict
that removing a protein from yeast with a greater $v_i$ harms more
biologically important pathways and would thereby be more likely to
destroy viability.  The percentage of proteins having a given $v_i$
which are essential ( (number of essential proteins of a given
$v_i$)/(number of proteins of the given $v_i$) ) is shown in
Fig.~\ref{fig:corr-ess-v}.  This percentage has strong positive
correlation with $v_i$, in agreement with the prediction.

What are the specific connectivity properties that produce a large
$v_i$ for a specific protein?  To a first order approximation, $v_i$
is proportional to the degree of connectivity of the $i^{\rm th}$
protein.  Since a protein with $k$ interactions is usually connected
to at least $p\cdot k$ proteins, in the first order $v_i$ is
proportional to $k_i$.  However, the protein interaction network
displays small world properties\footnote{The graph diameter (the
  maximum amongst all the shortest paths between all pairs of
  vertices) of the protein interaction network is 12. The average path
  length of the path between any two proteins is 4.23.}, Therefore,
the correction to $v_i$ from higher order connections should be
included.  For example, if the number of next-nearest neighbors of a
protein is much greater than the number of nearest neighbors, then the
contribution from the next-nearest neighbors is comparable to that of
the nearest neighbors.  In such a case, the proteins with the same
$k_i$ have a broad distribution of $v_i$ as in our results.  The value
of $v_i$ gives more extensive information about the protein's
connectivity in the network beyond that of its nearest neighbors.

Our method is advantageous because we can identify important proteins
that might otherwise not be considered significant because they have
lower first-order interaction degree.  Such proteins probably control
other essential proteins through a few critical interactions.  To
illustrate the power of this approach compared to merely counting the
nearest neighbor degree of interactions, we rank the proteins by $v_i$
and compare the result to the ranking by $k_i$ (see
Table~\ref{tab:compare}).  For example, 61\% of the proteins in the
top 2\% of $v_i$ are essential, whereas only 52\% of the proteins in
the top 2\% of $k_i$ are required for viability.  Such a result
suggests the essential proteins with higher $v_i$ not only have more
interactions but are also more likely to interact more frequently with
other proteins, which also tend to be essential.  A similar
observation has been reported by Gavin, {\it et al.}~\cite{Gavin}, and
our independent evidence supports their hypothesis.

\begin{table}[htbp]
  \begin{tabular}{|c||c|c|c|}
    \hline
    All Proteins & \multicolumn{3}{l}{Essential Proteins}\vline\\
    \hline
    \hline
    Percentile & by $v_i$ & by $k_i$ & by $v_i$ (randomize)\\
    \hline
    2\%(94) & 61\% & 52\% & 53\% \\
    5\%(234) & 53\% & 47\% &  50\% \\
    10\%(469) & 48\% & 46\% & 48\% \\
    25\%(1173) &39\% & 38\% & 38\% \\
    \hline
  \end{tabular}

  \caption{The percentage of essential proteins in
    selected percentiles ranked by $v_i$ and the degree of connection
    $k_i$.  In the top 92 proteins ranked by $v_i$, 61\% of them
    are essential while only 52\% of essential proteins are captured when
    ranked by $k_i$.  The third column is a control in which the $v_i$ are
    recalculated for a (quasi-)randomized graph in which edges have
    been swapped while retaining the degrees of connection of all vertices in
    the original graph. Identifying essential proteins by calculating 
    $v_i$ performs consistently better than only computing $k_i$, 
    demonstrating the significance of nonlocal structure beyond
    that of nearest neighbor relations.  If we randomly perturb the 
    global graph structure, the ability to identify essential proteins 
    drops, even though the degree of connection at each vertex is unchanged.}
  \label{tab:compare}
\end{table}

The proteins with 10 highest $v_i$ are listed in
Table~\ref{tab:pList1}.  The full list of proteins with their $v_i$
can be found in the supplemental web site\footnote{\tt
http://www.nas.nasa.gov/Groups/SciTech/nano/msamanta/projects/percolation/index.php}.
A selection of a few essential proteins with high $v_i$ but low $k_i$ is
also shown in Table~\ref{tab:pList2}.

\begin{table}[htbp]
  \begin{tabular}{|c|c|c|c|}
    \hline
    protein & $v_i$ & $k_i$ & viability \\
    \hline
    \hline
      SRP1 & 0.0623 & 196 & inviable \\
      TEM1 & 0.0531 & 115 & inviable \\
      JSN1 & 0.0524 & 282 & viable \\
      YDL213C & 0.0516 & 58 &  viable\\
      CKA1 & 0.0513 & 65 & viable \\
      NUP116 & 0.0505 & 146 & inviable \\
      ERB1 & 0.0494 & 55 & inviable \\
      HHF1 & 0.0486 & 74 & viable \\
      NOP2 & 0.0479 & 48 & inviable \\
      CDC95 & 0.0475 & 48 & viable\\
    \hline
  \end{tabular}
  \caption{List of the proteins with 10 highest $v_i$. }
  \label{tab:pList1}
\end{table}

\begin{table}[htbp]
  \begin{tabular}{|c|c|c||c|c|c|}
    \hline
    $k_i$ & protein & $v_i$ & $k_i$ & protein & $v_i$ \\
    \hline
    \hline
      & UTP8 & 0.0084      &    & MAK11 & 0.0127 \\ 
      & YKL088W & 0.0081   &    & BMS1 & 0.0124  \\  
    3 & DYS1 & 0.0075      &  5 & YPR144C & 0.0117 \\
      & TRL1 & 0.0070      &    & ACS2 & 0.0113   \\
      & GRS1 & 0.0068      &    & DIP2 & 0.0112   \\
    \hline
      & RLP24 & 0.0115     &    & NOP14 & 0.0133  \\ 
      & ROK1 & 0.0106      &    & NOC3  & 0.0131   \\
    4 & SPB4 & 0.0101      &  6 & SEN1 & 0.0124   \\
      & MES1 & 0.0094      &    & YLL034C &0.0123  \\
      & SEC18 & 0.00868    &    & DIB1 & 0.0110   \\
    \hline
  \end{tabular}
  \caption{A selection of a few essential proteins with
high $v_i$ but low $k_i$.}
  \label{tab:pList2}
\end{table}

\subsection{Distribution of $\beta_{ij}$}
\begin{figure}[htbp]
  \includegraphics[width=3in]{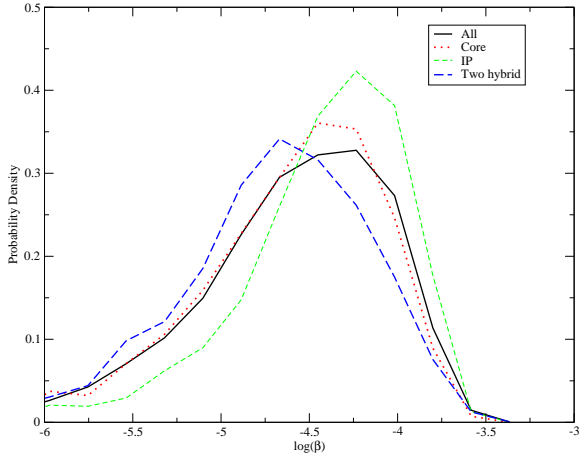}
  \caption{Normalized distributions of $\log(\beta_{ij})$ for
    different subsets of interactions. The solid line represents the
    distribution for all interactions in the data.  The dotted line
    corresponds to the core set extracted by Deane, {\it et
      al}\cite{Deane}.  The short dashed line refers to interactions
    obtained by immunoprecipitation, and the long dashed line
    represents the subset of interactions derived from high-throughput
    two-hybrid tests.}
  \label{fig:h_beta}
\end{figure}

The interactions in the network can be grouped by the experimental
methods used to detect them.  We score each interaction within the
network by $\beta_{ij}$.  The distribution of
$\log(\beta_{ij})$(Fig.~\ref{fig:h_beta}) provides a mechanism to
detect differences amongst different subsets of interactions obtained
by varied experimental methods.  In Fig.~\ref{fig:h_beta}, we compare
the distribution of $\log(\beta_{ij})$ from the whole network to
distribution derived from several subsets of the network.  First, we
use the subset, as the core set, of the interactions that was derived
by Deane {\it et al.}~\cite{Deane}.  Interactions in the core set are
statistically verified to reduce the false positive rate, yielding
1925 interactions (excluding self-interacting pairs).  The
distribution of $\log(\beta_{ij})$ for the core set is similar to that
obtained for the entire network.  However, upon comparing the
distribution of $\log(\beta_{ij})$ for subsets of those interactions
obtained from different experimental procedures, differences emerge.
For example, interactions measured by immunoprecipitation tends to
have a larger $\beta_{ij}$, so that the distribution of
$\log(\beta_{ij})$ of this subset shifts to the right.  In contrast,
the distribution for the subset of interactions measured with
high-throughput two-hybrid tests display the opposite trend.

If $e_{ij}$ is the only edge linking two clusters, the contribution of
a particular realization of the percolation procedure to $\beta_{ij}$
is proportional to the product of the sizes of the two clusters.  Hence,
an edge with a greater $\beta_{ij}$ has a greater tendency to link two
large modules or clusters in the network.  With this notion in mind,
an examination of Fig.~\ref{fig:h_beta} suggests that the IP method is
possibly more sensitive to interactions between proteins in different
large modules while the two-hybrid tests are better suited to
detecting interactions which tend not to link larger modules.

The discrepancy between the IP method and the two-hybrid tests might
reflect the underlying biochemical differences between the two
methods.  Unlike IP, the two-hybrid test is an {\it in vivo}
technique, and thus it can detect transient and unstable
interactions\cite{Mering}.  Our analysis of the distribution of
$\log(\beta_{ij})$ for the two-hybrid data is a quantitative
demonstration that these transient and unstable interactions
contribute less to the integrity of the interaction network.

\section{Conclusion}
\label{sec:con}

We presented a stochastic algorithm that explored the global
connectivity properties of a protein interaction network.  This
percolation-based algorithm allowed us to assign weights to vertices and
edges according to non-local topological properties.  We applied the
algorithm to the protein interaction network for yeast and found that
the percentage of essential proteins correlated strongly with $v_i$.
Importantly, the values of $v_i$, which incorporated the knowledge of
connections beyond the nearest neighbors, could more successfully
discriminate essential proteins than a method based solely on local
connections.  In addition, the essential proteins with greater $v_i$
not only possessed more interactions with any other proteins but also
displayed more interactions with other {\em essential} proteins.  This
result suggested that essential proteins along with other proteins
having greater $v_i$ might form a ``core network'' with a higher
density of interactions within the ``core network'' than the
background network.  If this unverified hypothesis is confirmed, then
we would gain significant insight into the evolution of a protein
interaction network.  Are the proteins in this ``core network'' in
general more evolutionarily conserved than others?  Hunter {\it et al.}
claimed that there is significant negative correlation between each
protein's degree of connectivity and protein evolutionary rate, and
that evolutionary change may occur largely by coevolution~\cite{Fraser}.
If this is indeed so, we expect a stronger correlation between $v_i$ and
protein evolutionary rate, since $v_i$ provides a better resolution
than the degree of connectivity for proteins' positions in their
interaction network.

The $\beta_{ij}$ scores for interaction could distinguish the differences
between different experimental methods for measuring protein
interactions.  Such a quantitative measure of the distinction amongst
the experimental approaches will aid the interpretation of the
proteomic data.

In principle, $c_{ij}$ can be calculated exactly given a percolation
probability $p$.  However, this would require recursive iterations
over all possible sub-graphs.  Our stochastic approach efficiently
obtains the approximations to the exact value of $c_{ij}$, $v_i$ and
$\beta_{ij}$.  In this work, we model the interaction network as a
static graph with uniform weight on each edge.  For a biological
system, dynamical aspects need to be incorporated.  Various
experimental methods for probing the physical interactions between
proteins respond differently to the dynamics of biological systems.
The two-hybrid test is more sensitive to transient interactions while
the IP method is more sensitive to large and stable protein complexes.
The differences might be addressed from different dynamics aspects in
the interaction network.

With regard to future pursuits, we note that it is also possible to
use $\beta_{ij}$ to cluster vertices within a random graph.  The
$\beta_{ij}$ score for a random graph is similar to the edge
``betweenness'', defined as the number of shortest paths between all
pairs of vertices passing through a given edge.  An edge with a
greater $\beta_{ij}$ is likely also an edge with a greater edge
``betweenness'', because such an edge has great tendency to bridge two
different clusters or modules.  Clustering utilizing edge
``betweenness'' have been successfully applied to certain types of
random networks\cite{Newman}.  We expect that results similar to those
shown in Fig.~\ref{fig:smallnet} could be achieved with $\beta_{ij}$
not only for this small test graph but more significantly for larger
graphs in which the computational cost of calculating edge
``betweenness'' is prohibitive.  For the present, however, the idea of
percolation on random networks provides a natural mechanism for
revealing dominant cluster structure within a graph.  We hope such
natural cluster structure will provide further details about the
protein interaction network.

\acknowledgements{ We thank Hao Li and Shoudan Liang for fruitful
  discussion.  C.~S.~Chin also likes to thank Yigal Nochomovitz for
  critical reading of the manuscript.  C.~S.~Chin is supported by Sandler
  Opportunity Grant.  M.~P.~Samanta is supported by NASA contract
  DTTS59-99-D-00437/A61812D to CSC.}

\end{document}